\documentclass[showkeys,superscriptaddress,showpacs,preprintnumbers,aps,twocolumn]{revtex4}
\usepackage[latin1]{inputenc}
\usepackage{epsfig}
\usepackage{amsfonts}
\usepackage{amssymb}

\newcommand{\ITP}{Institut f{\"u}r Theoretische Physik, 
  Technische Universit{\"a}t Berlin,
  Hardenbergstra{\ss}e 36, 10623 Berlin, Germany}

\newcommand{\GOETTINGEN}{Drittes Physikalisches  Institut, 
  Universit\"at G\"ottingen,
  B\"urgerstra\ss e 42-44, 37073 G\"ottingen, Germany}

\begin{document}

\date{July 2, 2003}

\preprint{Accepted for publication in Phys. Rev. Lett. (2003)}
\title{A hybrid model for chaotic front dynamics: From semiconductors to
  water tanks}
\author{A. Amann}
\affiliation{\ITP}
\author{K. Peters}
\affiliation{\GOETTINGEN}
\author{U. Parlitz}
\affiliation{\GOETTINGEN}
\author{A. Wacker}
\affiliation{\ITP}
\author{E. Sch\"oll}
\affiliation{\ITP}

\begin{abstract}
  We present a general method for studying front propagation in nonlinear systems with a global
  constraint in the language of hybrid tank models. The method is
  illustrated in the case of semiconductor superlattices, where the
  dynamics of the electron accumulation and depletion fronts shows 
  complex spatio-temporal patterns, including chaos. We show that this
  behavior may be
  elegantly explained by a tank model, for which analytical results
  on the emergence of chaos are available. In particular, for the case of
  three tanks the bifurcation scenario is characterized by a modified
  version of the one--dimensional iterated tent--map.
\end{abstract}


\maketitle

\paragraph*{Introduction.}

Moving fronts are the source of complex patterns in very different
areas of physics \cite{CRO93}, chemistry \cite{KAP95a,MIK94} and even
biology \cite{SHE97,MUE02}.  While the propagation of single fronts is
well understood for many systems, only little is known on how front
generation and annihilation processes or collisions between fronts
influence the possible bifurcation scenarios. 

\begin{figure}[tbp]
  \centering
  \epsfig{file=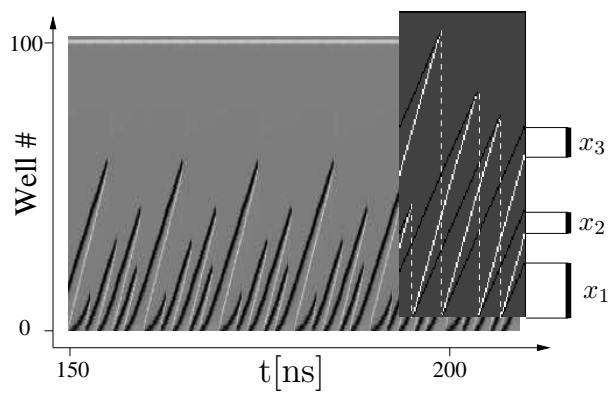,width=\columnwidth}

  \caption{(a) Space-time plot of the electron density evolution in a semiconductor 
superlattice. Electron accumulation and depletion fronts are indicated by light and
dark shading, respectively. The full microscopic sequential tunneling  model of 
Ref.  \onlinecite{AMA02a} for $N=$100 quantum wells and
contact conductivity $\sigma = 0.5 \quad\Omega^{-1}m^{-1}$ is used for a bias
$U =0.95$ V.
  The inset shows a simplified view of the pattern, which is adopted
  in the front and the tank model. The dashed lines denote the switching times $t_m$.
} 
  \label{fig:front_microscopic}
\end{figure}

For the sake of concreteness we illustrate our method with a semiconductor superlattice
system, which is well studied experimentally and theoretically
\cite{ESA74,KAS95,HOF96,PAT98,WAC02,BON02} and 
shows a wide range of
complex front patterns (cf. Fig.~\ref{fig:front_microscopic})
including chaos \cite{BUL95,LUO98b,AMA02a}. 
Similar front patterns occur in many other systems \cite{SCH01}, 
e.g., spatially continuous
semiconductor models describing impurity impact ionization breakdown \cite{CAN01} or
globally coupled heterogeneous catalytic reactions \cite{GRA93a}.
Our aim is to point out a striking connection of those front patterns with discrete hybrid
models of  $n$ tanks such as they occur, for instance, in industrial production processes 
\cite{CHA93a},
and derive a simple description of the front generation and annihilation dynamics in terms of an 
$n$-tank model which allows for an analytical treatment of the bifurcation scenarios.
In particular, an unusual chaotic scenario will be characterized  
by this generic model.  We stress that the presented methods 
do not depend on the pecularities of the superlattice model, and are
expected to be equally applicable to other front systems with global
coupling.

\begin{figure}[tbp]
  \centering
  \epsfig{file=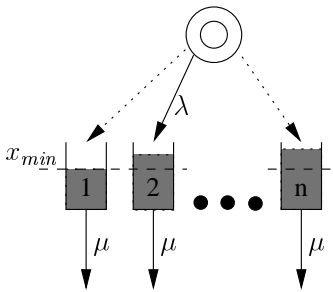,width=0.5\columnwidth}
  \caption{Scheme of an $n$-tank switched arrival system with minimal filling
  height $x_{min}$.}
  \label{fig:arrival}
\end{figure}

In order to formulate a generic model for front propagation, we will first
develop a {\em front model} based on the spatio-temporal dynamics
in a semiconductor superlattice.
The dynamical variables are the positions of the accumulation
and depletion fronts which evolve with a velocity determined by the number of
fronts present in the system.  The number of fronts changes according
to simple rules which mimic generation of fronts at the emitter and
the annihilation of fronts. 
Such models are quite general and similar ideas have been sketched
for front propagation in bulk GaAs  \cite{CAN01}.
It will be shown that this front model is identical to 
an {\em $n$-tank switched arrival system} \cite{CHA93a,PET03}, if one identifies
the spacing $x_i$  (see Fig.~\ref{fig:front_microscopic})
between the leading depletion and the trailing accumulation front 
with the filling heights of the tanks.
The tanks drain at a constant rate $\mu$, while one of the tanks 
is connected to a server and filled at a rate $\lambda = \tilde{n} \mu$
(Fig~\ref{fig:arrival}), where $\tilde{n}$ is the number of tanks which 
are not empty. In the present model the server will switch
to a different tank, if this tank is empty and 
the filling level of the currently
filled tank is larger than $x_{min}$.
In the simplest case $n=3$, the
tank--model transforms into a one--dimensional iterated map which is very
similar to the well known tent--map.  We will show that the front
dynamics in semiconductor superlattices as simulated in
\cite{AMA02a} can largely be explained by such a
one--dimensional map. In particular, atypical border collision
bifurcations occur \cite{BAN99}. 
The general type of models we consider here is called {\em hybrid}
since they contain continuous as well as discrete state variables.

\paragraph*{Modeling the  front dynamics.}

\begin{figure}[tbp]
  \begin{center}
    \epsfig{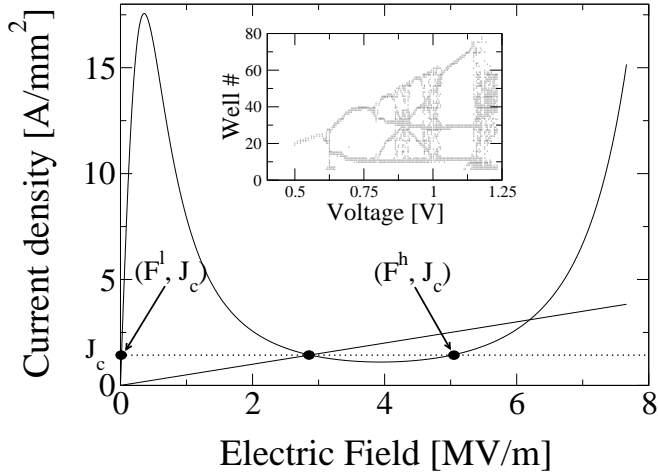}
    \caption{Current density vs electric field
      characteristic at the emitter barrier (straight line, Ohmic conductivity
      $\sigma = 0.5 \quad\Omega^{-1} m^{-1}$) and
      between two neutral wells (N-shaped curve) for the 
      microscopic superlattice model \cite{AMA02a}.  $J_c$ denotes the intersection
      point of the two characteristics. $F^l$ and $F^h$ are the stable
      low and high field states, respectively, for the current density $J_c$. 
      Inset: Positions where
      accumulation and depletion fronts annihilate vs. voltage.  }    
    \label{fig:homogen}
  \end{center}
\end{figure}

The superlattice dynamics is described by the electron densities 
$n_m(t)$ in the quantum wells, labeled by $m=1,2\ldots N$.
The current density $J_{m \to m+1}(n_m, n_{m+1}, F_m)$ between adjacent
wells is evaluated by the sequential tunneling model, where $F_m<0$
is the electric field between these two wells \cite{WAC02}. 
A microscopic calculation of $J_{m \to m+1}$ 
yields an N--shaped current density vs.
electric field characteristic as depicted in
Fig.~\ref{fig:homogen}.  In the electrically neutral regions of the superlattice,
there are two stable states for fixed current density $J$, one at a
low field $F^l(J)$ and  one at a high field $F^h(J)$
(Fig~\ref{fig:homogen}). Since in the following 
$J$ is varied only in a small range, we make the approximation
that both fields do not depend on the current density, and set $F^l(J) = 0$,
$F^h(J) = F^h>0$. 

The emitter boundary condition (at well $m=1$) is characterized
by its intersection point $J_c$ with the characteristics
between neutral wells.
For small $J_c$ propagating  inhomogeneous field 
distributions alternating between $F^l$ and $F^h$ occur, 
see Fig.~(\ref{fig:front_microscopic}). 
For a transition from a {\em low field domain} $F_m = 0$ to a {\em high field
domain} $F_{n} = -F^h$ $ (n>m)$ a negatively charged
electron accumulation front located at the domain boundary,
 is required by Poisson's equation.  Let us
denote by $a_i$ the distance of its center of charge from the emitter.
The index $i = 1 \ldots N_a$ labels different accumulation fronts such
that $a_i < a_{i+1}$, where  $N_a$ is the total number of accumulation fronts
in the system.
Similarly, a transition from a high field domain to a low field domain
can be attributed to an electron depletion front whose center of
charge is located at $d_i$ with $i = 1 \ldots N_d$, and $N_d$ the
number of depletion fronts. Since we only allow for alternating
accumulation and depletion fronts, we have $N_d - N_a \in \{ -1,0,1\}$.

The applied voltage $U$ between emitter and collector defines the
partial length $L_h=U/F^h$ of the 
superlattice which is at high field. This imposes a
global constraint on the front positions by
\begin{equation}
  \label{eq:voltage}
  L_h =  \sum_{i=1}^{N_d} d_i - \sum_{i=1}^{N_a} a_i \quad \text{mod } L.
\end{equation}
Here  $L$ denotes the total length of the superlattice. The
expression $\text{mod } L $ in  (\ref{eq:voltage}) means that $L$ has to be
added if $a_{N_a}>d_{N_d}$ such that $L_h \in [0,L]$. 

The front velocities are  given functions of the total current
$\dot{a}_i =  v_a(J)$, $\dot{d}_i= v_d(J)$,
as shown in Refs.~\cite{CAR00,AMA01}. 
Differentiating eq.~(\ref{eq:voltage}) yields 
$v_d(J)/v_a(J)=N_a/N_d$. This determines the current
$J(N_a/N_d)$, which is a monotonically increasing function, 
as $v_d(J)$ and $v_a(J)$ are increasing and decreasing, respectively
\cite{AMA02a}.
Rescaling the time such that $v_a + v_d  = 2$ gives
\begin{equation}
  v_a = \frac{2 N_d}{N_a + N_d}\, \quad
  v_d = \frac{2 N_a}{N_a + N_d} 
  \label{eq:velocities}
 \end{equation}

Front injection at the
emitter is governed by $J_c$. For $J < J_c$ the
region close to the emitter is pinned at a low field \cite{WAC02}.
Hence an accumulation front is injected, if the preceding front is a
depletion front. Due to the finite width and buildup time of the
fronts new fronts can not be injected arbitrarily fast.
Phenomenologically we therefore introduce a distance parameter $p_h$,
and assume that front generation is suppressed while $d_1 < p_h$,
where $d_1$ is the position of the first depletion front.  In
the same fashion, a depletion front is injected if $J$ rises above
$J_c$ and $p_l < a_1$, where $p_l$ is a parameter describing the
minimum front distance for depletion front injection.
As the current $J$ is a monotonic function of $N_a/N_d$, the
condition $J \gtrless J_c$ can be transformed into
$N_a / N_d \gtrless r_c$ with a new critical parameter $r_c$.

The processes which reduce the number of fronts are collisions of two
fronts of opposite polarity and the running out of fronts at the
collector. Both processes affect $N_a/N_d$ and potentially trigger generation 
of a new front at the emitter. 

Let us summarize the rules for front dynamics:
\begin{enumerate}
\item[FI] The positions of the accumulation fronts 
$a_i$ for $i = 1 \ldots  N_a$ and depletion fronts
$d_i$ for $i = 1 \ldots  N_d$ evolve according to 
$\dot{a}_i=v_a$ and $\dot{d}_i=v_d$
with the velocities (\ref{eq:velocities})
until one of the following rules applies.
\item[FII] If $N_a / N_d < r_c$ and $p_h < d_1 < a_1$ then increase $N_a$ by
one, re-index $a_i \to a_{i+1}$ for all $i$ and set $a_1 = 0$ (injection of accumulation front).
\item[FIII] If $N_a / N_d > r_c$ and $p_l < a_1 < d_1$ then increase $N_d$ by
one, re-index $d_i \to d_{i+1}$ for all $i$ and set $d_1 = 0$ (injection of depletion front).
\item[FIV] If $a_{i'} = d_{j'}$ for any $i',j'$  then decrease $N_a$ and
$N_d$ by one, re-index $a_{i+1}\to a_i$ for $i \geq i'$ and $d_{j+1}\to
d_j$ for $j \geq j'$ (annihilation of front pair). 
\item[FV] If $a_{N_a}> L$ decrease $N_a$ by one (accumulation front hits collector).
\item[FVI] If $d_{N_d}> L$ decrease $N_d$ by one (depletion front hits collector).
\end{enumerate}

Note that the voltage parameter $L_h$ only enters in the initial 
condition for the front positions (see Eq. (\ref{eq:voltage})). 

In the following we will restrict ourselves to the case $r_c < 1$ and
$p_l=0$. 
Rule FII does not apply as long as $N_a > r_c/(1-r_c)$ and $N_a$ can then only
decrease. Consequently
$r_c$ imposes the following limits on the number of fronts:
\begin{eqnarray}
  \label{eq:domain_constrained}
  N_d \leq  n; \quad N_a &\leq& n-1;
\end{eqnarray}
where $n$ is the largest integer less than $1/(1-r_c) + 1$.
The injection of an accumulation front is immediately
followed by the injection of a depletion front (FIII). This
detaches a high field domain from the emitter and opens a new one.
Furthermore we restrict ourselves to the case where no fronts reach the collector 
(i.e. $L > n L_h$). This corresponds to the situation in
Fig.~(\ref{fig:front_microscopic}) and we have $N_a = N_d -1$.

\paragraph*{The n-tank  model.}

Under the restrictions given above we may choose the lengths of the high
field domains (cf. inset of Fig.1) $x_1=d_1$ and
$x_{i} = d_{i} - a_{i-1}$ for  $i = 2 \ldots N_d$
as new dynamical variables governed by the rules:
\begin{enumerate}
\item[TI] The lengths evolve according to
$\dot{x}_i=-\mu+\lambda\delta_{i1}$ with
$\mu=2/(2 N_d-1)$ and $\lambda=N_d \mu$.
\item[TII] If $N_d < n$ 
and $x_1 > p_h$ then increase $N_d$ by one,
re-index $x_i \to x_{i+1}$ for all $i$ and set $x_1=0$.
\item[TIII] If $x_{i'} = 0$ then decrease $N_d$ by one, re-index $x_{i+1}
  \to x_{i}$ for all $i\geq i'$.
\end{enumerate}
The voltage constraint (\ref{eq:voltage}) is expressed in the new
variables by  $\sum_k x_k=L_h$. 
 
It is very illustrative to interpret this model in the language of an
$n$-tank switched arrival system (cf. Fig.\ref{fig:arrival}). A domain
may be viewed as a tank with fluid content $x_i$. One tank
is filled at rate $(N_d-1)\mu$, while $N_d-1$ tanks
are drained at rate $\mu$, and $n-N_d$ tanks are empty and
waiting for filling, because
the currently filled tank has not yet reached the minimum filling
level $x_{min}=p_h$.

\begin{figure}[tbp]
\epsfig{file=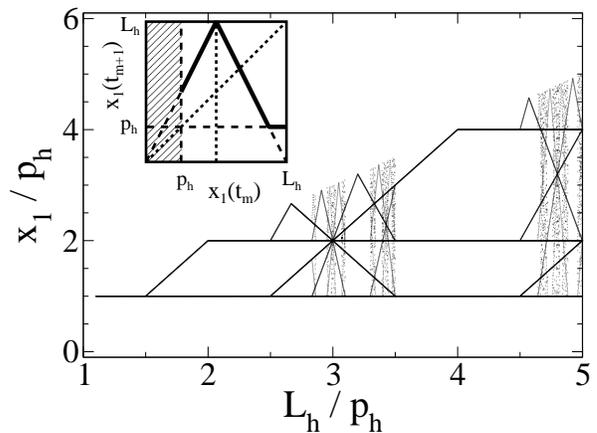,width=0.9\columnwidth}
\caption{ Bifurcation diagram of $x_1$ vs. $L_h$ in units of $p_h$. Inset:
Graph of the return  map for the $n=3$ tank
 model  Eq. (\ref{eq:flat_bottom_tent}). In the shaded region the map is  not defined. 
}
\label{fig:tent_map}
\end{figure}

Let us  first  discuss the limiting case $p_l=p_h=0$. If a tank is empty
at time $t_m$ (i.e. $x_k(t_m)=0$), the rules TII and TIII apply
instantaneously and the system starts to fill this tank
($x_1(t_m^+)=0$). In this case no tank
has to wait for filling. Apart from the re-indexing procedures we
obtain a switched arrival system \cite{CHA93a}. As shown in
\cite{SCH95h} the system is chaotic for all $n > 2$ and has a constant
invariant probability measure.
By sampling the dynamics at the switching times $t_m$ we obtain the  {\em
  Poincar\'e map} $P:\vec{x}(t_m)\mapsto\vec{x}(t_{m+1})$: 
\begin{equation}
  \label{eq:Poinc_map1}
\vec{x}(t_{m+1})=\vec{x}(t_m^+)+\dot{\vec{x}}\Delta t_m
\end{equation}
where $\dot{\vec{x}}=(\lambda-\mu, -\mu,\ldots,-\mu)$ is a constant vector
and $\Delta t_m=t_{m+1}-t_m=\min_{j\neq 1}(x_j(t_m^+)/\mu)$. 
Generally this is an $n$ dimensional mapping which acts due to the
normalization $\sum_ix_i=L_h$ on the $n-2$ dimensional boundary of a
regular $n$-simplex embedded in $\mathbb{R}^{n}$. 
For $n=3$ we can reduce this map to a one-dimensional map of the
interval $[0,L_h]$ onto itself:
From the rules we obtain
$x_1(t_m^+)=0$, $x_2(t_m^+)=x_1(t_m)$, and $x_3(t_m^+)=L_h-x_1(t_m)$.
If $x_1(t_m)< L_h/2$, the tank $2$ empties after 
$\Delta t_m=x_1(t_m)/\mu$ and we find $x_1(t_{m+1})=2x_1(t_m)$.
Otherwise, if $x_1(t_m)> L_h/2$, the tank $3$  empties after 
$\Delta t_m=(L_h-x_1(t_m))/\mu$ and we find $x_1(t_{m+1})=2(L_h-x_1(t_m))$.
Thus the dynamics is that of the well known {\em tent
  map}: $x_1(t_{m+1})=L_h-2|x_1(t_m)-L_h/2|$.

Now we consider the case $n=3$ with a finite minimum filling level
$p_h>0$.  Since by TII $x_1$ must be larger than $p_h$ for
switching, the Poincar\'e map is now only defined on the interval
$[p_h,L_h]$. For $x_1 \in [p_h,L_h-p_h/2]$ the map remains unchanged
from the tent map discussed before. If the switch to an empty tank
occurs at $x_1(t_m) > L_h - p_h/2$, the third tank will be empty
before the newly filled tank has reached the minimum filling level
$p_h$. Switching is therefore suppressed until $x_1(t_{m+1})=p_h$.
Thus for $n=3$ we obtain (cf. inset of Fig.\ref{fig:tent_map}):
\begin{equation}
x_1(t_{m+1})=\mathrm{Max}\left\{(L_h-2|x_1(t_m)-L_h/2|),p_h\right\}
\label{eq:flat_bottom_tent}
\end{equation}
The bifurcation structure of this map with respect to $L_h$ is
unusual (Fig.\ref{fig:tent_map}) but has similarities to the
flat-topped tent map \cite{GLA94,WAG02}.  Due to the horizontal
segments the map cannot show chaotic behaviour although it undergoes
several bifurcation cascades. A trajectory on the tent map segments
would explore the attractor of the tent map and will therefore either
fall on one of the flat segments or stay at an unstable periodic orbit
contained in the remaining part of the tent map.  
Once arrived at one of the flat segments the trajectory
continues on a stable periodic orbit. 
By analyzing the $k$-th 
return map of the modified tent map we
find that windows of period $k$ are found inside the intervals
$L_h/p_h\in (\frac{2^k-1}{4j-2},\frac{2^k+1}{4j-2})$ for 
$j\in I_k\subset \{j\in \mathbb{N} | j\leq 2^{k-2}-1\}$. Thus higher and higher stable periodic orbits
are arising in smaller and smaller intervals. 

\paragraph*{Discussion.}
In order to compare the bifurcation diagram of the modified tent map
(Fig.\ref{fig:tent_map}) with the bifurcation diagram of the
superlattice model (inset of Fig.~\ref{fig:homogen}), we note that the
voltage $U$ and the position of front annihilation in the superlattice
model correspond to the parameters $L_h$ and $x_1(t_m)$ in the
map, respectively.  We can therefore identify the periodic windows
found in the Poincar\'e map with the periodic pattern observed in the
front model. We note that the bifurcation scenarios agree very
well up to voltages of about $1.1$V, where a period-3 window occurs
(inset of Fig.~\ref{fig:homogen}). In particular, the remarkable
spider-like structure at $0.9$V is found for $L_h/p_h=3$ in the
bifurcation diagram of the map. The reason for this spider is
that the contracting flat region is mapped onto the unstable fixed
point of the map.  Further prominent features in both bifurcation
diagrams are the horizontal lines given by the first, and higher
iterated images of the minimal filling height.
Thus the character of the bifurcation scenario
of complex front dynamics is explained by considering a
one-dimensional iterated map with only one free parameter $L_h/p_h$.

The minimum filling height corresponds to the minimum 
domain width. For a real superlattice
system this quantity will slightly depend on the history
of the dynamical evolution. 
Therefore  the slope of the flat segments is not exactly
zero but attains a small finite value $\alpha$. In
this case a period-$k$ orbit is unstable if $|2^k \alpha| >1$. Close to
the spider-like points where periodic orbits with large period
$k$ occur,  we can then find a finite chaotic regime, as observed in our 
simulations with the microscopic model.
 
\paragraph*{Conclusions.}
\label{sec:conclusions}
We have introduced a hybrid tank model as a
simplified model for front dynamics in nonlinear systems with a global constraint.
This simplified model provides an explanation for the complex accumulation and
depletion front dynamics in semiconductor superlattices. 
In this unexpected analogy, the tank filling levels correspond to the widths of the high field
domains in the superlattice, while the position of the server determines which high field 
domain is connected to the emitter.  
Although this analysis has started from a superlattice model, 
the presented methods are very general
 and are expected to work successfully for other systems which provide
 a global coupling similar to Eq. (\ref{eq:voltage}).

\paragraph*{Acknowledgments.}
\label{sec:acknowledgments}
This work was partially supported by DFG in the framework of Sfb 555
and the Volkswagen Foundation (Grant No. I/76 279 - 280). 
We thank N. Janson, A.~Pikovsky and J. Schlesner for stimulating discussions.

\end{document}